 \documentclass[final,3p,times]{elsarticle}
\usepackage{amsmath,amssymb,amsthm,eucal}
\usepackage[american]{babel}
\usepackage{epsfig}
\usepackage{psfrag}
\usepackage{subfigure}
\usepackage{color}
\usepackage{soul}
\usepackage{bbm}
\usepackage{amssymb}
\usepackage{color}

\journal{Applied Mathematics Letters}

\newcommand{\BE}{\begin{eqnarray}}
\newcommand{\EE}{\end{eqnarray}}
\newcommand{\BD}{\begin{description}}
\newcommand{\ED}{\end{description}}

\def\q{\mathbf{q}}

\def\vv{\mathbf{v}}

\def\D{\mathbb{D}}

\begin{document}

\begin{frontmatter}



\title{A Multiphase First Order Model for \\ Non-Equilibrium Sand Erosion, Transport and Sedimentation}

\author{L. Preziosi}
\address{Department of Mathematical Sciences -- Politecnico di Torino \\ Corso degli Abruzzi 24, 10129, Torino, Italy.}
\author{D. Fransos}
\address{Optiflow Company \\ 27 Boulevard Charles Moretti, F-13014 Marseille, France}
\author{L. Bruno}
\address{Department of  Architecture and Design -- Politecnico di Torino \\ Corso degli Abruzzi 24, 10129, Torino, Italy.}


\begin{abstract}
Three phenomena are involved in sand movement: erosion, wind transport, and sedimentation. This paper presents a comprehensive easy-to-use multiphase model that include all three aspects with a particular attention to situations in which erosion due to wind shear and sedimentation due to gravity are not in equilibrium. The interest is related to the fact that these are the situations leading to a change of profile of the sand bed.
\end{abstract}


\end{frontmatter}


\section{Introduction}
When the shear stress exerted by wind on a sandy surface is sufficietly strong, sand grains are lifted from the sand bed and are transported by wind to sediment downstream. 
The raising sand grains follow a ballistic trajectory  influenced by drag and gravity, eventually impacting again on the surface and inducing new particles to detach from the surface. This phenomenon, knows as saltation, generates a layer close to the sand bed with a typical maximum height of 10-20 cm.
Saltation is the main reason of erosion of sandy surfaces and together with the consequent sedimentation of sand particles it is the main reason of dune motion and accumulation of sand in specific regions where recirculation occurs. 

\begin{figure}[h!]
\begin{center}
\includegraphics[width=1\textwidth]{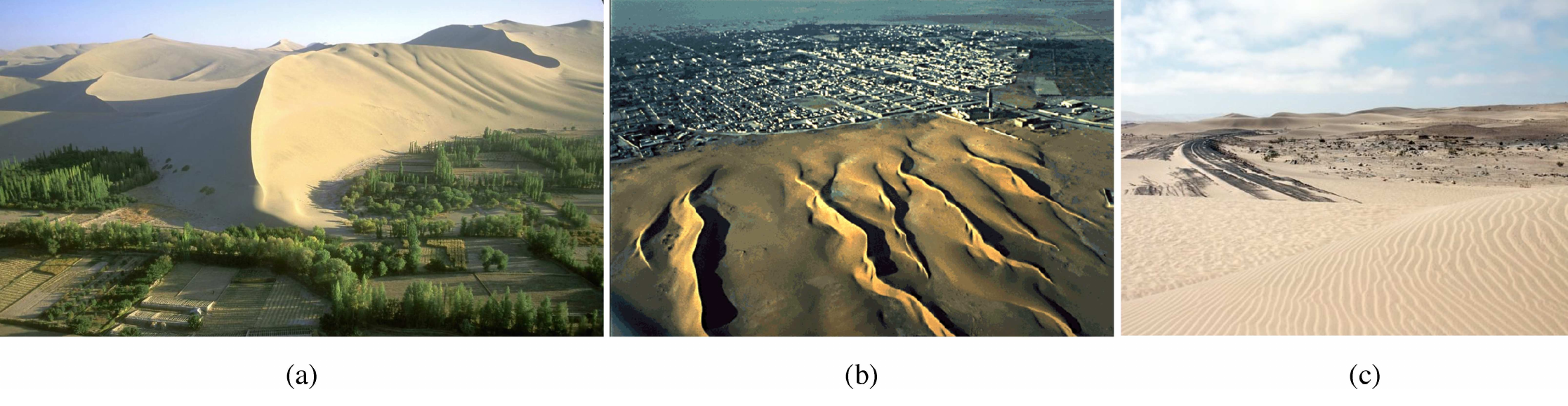}
\end{center}
\vspace{-.8cm}
\caption{Windblown sand interaction with anthropic activities: farmlands (a), towns (b), railways (c)}
\label{figure_1}
\end{figure}

The engineering interest in understanding and simulating the dynamics of  windblown sand, e.g. dune fields of loose sand, is dictated by their interaction with a number of human infrastructures in arid environments,
such as roads and railways, pipelines, industrial facilities, farmlands, towns and buildings as shown in Fig. \ref{figure_1} \cite{Alghamdi_2005}.

Moving intruder sand dunes, soil erosion and/or sand contamination can be comprehensively ascribed, from a phenomenological point of view, to non-equilibrium conditions, where the two processes, erosion and sedimentation, do not balance, leading to the erosion or deposition of sand on the soil and eventually to the evolution of that interface. In other terms, such non-equilibrium situations are the most interesting cases from the applicative point of view.

Bagnold \cite{Bagnold_1941} 
was the first who studied sand erosion and postulated a relation for the sand flux, determining the importance of wind speed and of the related shear stress on the sand surface. Later authors 
%(Andreotti et al. 2002; Gillette and Walker, 1977; Ji et al. 2004; Kerr and Nigra, 1952; Nalpanis et. al. 1993; Pie and Tsoar, 1990; Sauermann et al., 2001; Shao and Li, 1999)
\cite{ andreotti_parte1, Gillette_1974, Ji, Kerr, Nalpanis, Tsoar, Sauermann01, Shao_1999}
introduced several corrections to Bagnold's rule, but all the models have in common the observation that a sand grain is ejected from a sand bed if and only if the shear stress at the surface is larger than a threshold value. 

Sauermann et al. \cite{Sauermann01} 
observed that saltation reaches a steady state after a transitory phase of 2 seconds. After this period the trajectories are statistically equivalent for the ensemble of grains. This phenomenon happens because the new ejected particles increase the sand concentration in the saltation layer and this reduces the speed of saltating grains. So, a steady state is reached when all particles are ejected with the same velocity 
%(Almeida et al., 2006, Anderson and Haff, 1991; Andreotti et al., 2010; Shao and Li, 1999).
(see also \cite{Almeida_2006,  Anderson_Haff_1991, Andreotti_2010, Shao_1999}). 
%
%\begin{figure}
%\centering
%\begin{center}
%\includegraphics[width=\columnwidth]{Saltation.eps}
%\end{center}
%\caption{Trajectories of sand particles. {\bf Metterla?}}
%\label{d_saltation}
%\end{figure}
%
A nice mathematical models of the saltation phenomenon is proposed by Herrmann and Sauermann \cite{Hermann00} 
who studied the dynamics of the surface of a dry granular bed dividing the sand bed into a non-moving time-dependent region providing sand mass and another time-dependent region above it in which sand particles can move transported by the wind. They propose a model averaged over the vertical coordinate, presenting a free boundary.

Ji et al. \cite{Ji} coupled a $k-\epsilon$ model with a multiphase approach in which 
the slip of the dispersed phase is modeled by an algebraic model.  
Similar turbulent one-dimensional models are proposed in \cite{Sauermann02, Parteli_2006, Sauermann03,  Sauermann01}, however without a multiphase coupling. 
Kang and coworkers 
%(Kang, 2012; Kang and Guo, 2006; Kang and Liu, 2010, Kang and Zou, 2011) 
\cite{Kang12, Kang10, Kang11} 
instead couple a multiphase model for the fluid flow with a particle method for the sand grains.
A similar coupling was also used in 
%(Almeida et al., 2006; Andreotti et al., 2010: Kok and Renno, 2009)
\cite{Almeida_2006, Andreotti, Kok} 
where however the wind flow was computed indipendently from the presence of sand particles via a suitable turbulence model, typically the k-$\epsilon$ model. 

Sedimentation has also been widely studied in the literature starting from several applications mainly in environmental and chemical engineering. One of the most important component in this phenomenon is the drag force experienced by the sedimenting particles that has driven a lot of attention by many authors as well reviewed in \cite{Barnea}.

%Farimani et al. \cite{Farimani} focused on the evolution of the free surface under the action of a two dimensional wind flow causing the erosion of the sand dune.

Differently from previous papers, here we will propose a comprehensive multiphase model for the entire process including sand erosion, wind transport, and sedimentation, that working also in non-equilibrium conditions is able to deal with the development of the stationary saltation layer starting from generic initial and boundary conditions and in particular from clear air and oversaturated situations. 
In order to do that we develop a so-called first order model (in time) of sand erosion, transport and deposition, that can be easily tuned using experimental test cases. The resulting advection-diffusion equation for the suspended phase can then be coupled with a $k-\omega$ model describing the turbulent fluid flow. The mathematical model can then be solved with the aid of the fundamental erosion/deposition boundary condition at the sand bed, that depends on the shear stress.

The plan of the paper is then the following. After this introduction, Section 2 presents the mathematical model mainly focusing on the advective phenomena, on the microscopic dynamics related to the collision between sand grains, and on the erosion boundary condition. The result of some numerical simulations focusing on how the stationary condition is reached when wind blows over a heterogeneous sand bed are reported in Section 3.
%
%%%%%%%%%%%%%%%%%%%%%%%%%%%%%%%%
\section{The Erosion/Transport/Deposition Model}
We consider the flow of sand as a multiphase system composed of sand grains in air. Single sand grains have a density $\hat\rho_s$ and float in air with a volume ratio $\phi_s$ (typically well below 1\%), so that the partial density of sand in air is $\rho_s=\hat\rho_s \phi_s$.
Saturation obviously implies that $\phi_f=1-\phi_s$ where $\phi_f$ is the volume ratio of air. 
The mixture of air and sand grains is flowing on a sandy surface having a close packing volume ratio $\bar\phi_s$.

Because wind flow is in a turbulent regime the fluid phase is modelled by the Reynolds-averaged Navier-Stokes equations
(RANS) equations. More precisely, a $k-\omega$ turbulence model is selected to provide the closure \cite{Menter}
\begin{equation}\label{NS}
\left\{
\begin{aligned}
&\nabla\cdot \vv_f=0\\
&\rho_f\left(\frac{\partial \vv_f}{\partial t}+\vv_f\cdot\nabla\vv_f
\right)= -\nabla {p}+\nabla\cdot[\rho_f(\nu_a+\nu_t)\nabla\vv_f]\\
&\frac{\partial k}{\partial t}+\nabla\cdot (k\vv_f)=
\nabla\cdot[(\nu_a+\nu_t)\nabla k]+P_k-\gamma\omega k\\
&\frac{\partial \omega}{\partial t}+\nabla\cdot (\omega\vv_f)=
\nabla\cdot[(\nu_a+\nu_t)\nabla \omega]+P_\omega-C_\omega\omega^2\\
\end{aligned}
\right.
\end{equation}
with standard boundary conditions. In (\ref{NS}) $k$ is the turbulent kinetic energy, $\omega$ is the specific dissipation rate, $\nu_a$ and $\nu_t$ are, respectively, air and turbulence viscosities, $P_k$ and $P_\omega$ are the production terms for $k$ and $\omega$, and $\gamma$ and $C_\omega$ are two empirical costants.

In describing the transport of sand we start observing that while sand particles are trasported by the wind they drift down with a characteristic sedimentation velocity due to the action of gravity \cite{Barnea}. In addition, particle collide giving rise to an extra-flux term 
$\q_{coll}$. Hence, one we can write the following equation for the sand volume ratio

\begin{equation}\label{mass}
\frac{\partial \phi_s}{\partial t}+\nabla\cdot(\phi_s\vv_s+\q_{coll})=0\,,
\end{equation}
where 
\begin{equation}\label{closure}
\vv_s=\vv_f-w_{sed}{\bf k}\,.
\end{equation}
This closure can be actually deduced under suitable modelling assumptions from a more general multiphase model involving mass and momentum balance for the suspended phase.

The sedimentation velocity $w_{sed}$ can be evaluated by the balance of drag and buoyancy forces and strongly depends on the grain size. For instance, if we define the particle Reynolds number as the one felt by the sand grains of diameter $d$ during their flow and therefore based on the relative velocity between air and solid particles, 
%\begin{equation}\label{Res}
$Re_s={\phi_f|\vv_f-\vv_s|d}/{\nu_f}$,
%\end{equation}
then
in the so-called Newton regime, corresponding to particle Reynolds numbers
above 500, the drag coefficient $C_D$ is approximately constant (for instance, $C_d=3$ is used in \cite{Parteli_2006, Sauermann01}), so that one has the classical relation 
\begin{equation}\label{vsed}
w_{sed}=\sqrt{\frac{4(\hat{\rho}_s-\hat{\rho}_f) g}{3\hat{\rho}_f C_d}d}\,.
\end{equation}
However, at the other extreme, i.e., for particle Reynolds number below few units, corresponding to the so-called Stokes regime, $C_d\approx \frac{24}{Re_s}$, so that one has the classical Stokes sedimentation velocity
\begin{equation}\label{vsed2}
w_{sed}=\frac{(\hat{\rho}_s-\hat{\rho}_f) g}{18\hat{\rho}_f \nu_f }d^2\,.
\end{equation}
\begin{figure}
\begin{center}
\includegraphics[width=.8\columnwidth]{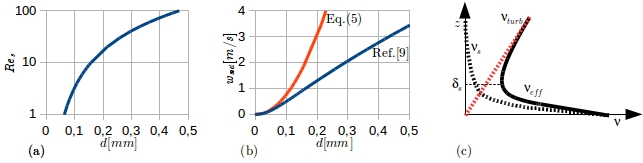}
\end{center}
\vspace{-.6cm}
\caption{(a) Particle Reynolds number as a function of the sand grain diameter of a sedimenting particle in air.
(b) Sedimentation velocity of sand grains in still air in the Stokes limit of Eq. (5) and according to the experiments summarize in \cite{Barnea}.
(c) Qualitative dependence of the diffusivity coefficient $\nu_{eff}$ from the distance from the sand bed.}
\label{figure_2}
\end{figure}
Considering that in aeolian sand trasport the phenomenon is limited to the first few centimeters from the ground and that there the particle Reynolds number is below 100 (see Fig. \ref{figure_2}a), a better evaluation of the sedimentation velocity with respect to Eqs. (\ref{vsed}, \ref{vsed2}) can be obtained fitting the experimental dependence of the drag coefficient on the particle Reynolds number as reviewed in \cite{Barnea} and shown in Fig. \ref{figure_2}b.

%{\bf Non so se citare in qualche modo le references successive}
%"Specifically, in their particle simulation model of sand transport Kang and coworkers
%\cite{Kang12, Kang10} implement Di Felice's drag law.
%As already adviced in other articles \cite{Almeida_2006,andreotti_parte1, Ji}
% %(Andreotti et al., 2002; Ji et al., 2004; Almeida et al., 2006) 
%this same approximation should not be used when interested in modelling the entire fluid-dynamic problem, as in our case."

Coming to the collision term  ${\bf q}_{coll}$, already introduced by Batchelor \cite{Batchelor83}, neclecting it would imply that sand grains are only transported under the action of drag and gravity. However, collisions among particles have the important non-negligible effect of generating a sort of diffusion of sand particles from higher to lower density areas, that results fundamental  in this modelling framework for the stationary formation of the saltation layer.

For high volume ratios near close packing, Auzerais et al. \cite{Auzerais}  suggested the following nonlinear law on the basis of experimental data
\begin{equation}\label{Fcoll2}
{\bf q}_{coll}=
-D_{eff}\nabla\frac{\phi_s^k}{\bar \phi_s-\phi_s}\,, \qquad{\rm with}\quad k\in [2,5]\,.
\end{equation}
Such a term enforces the need of avoiding that the close packing volume ratio $\bar \phi_s$    is reached for the sedimenting mass. 
However, as $\phi_s\ll\bar \phi_s$ in wind-blown sand applications, the relation can be simplified to
\begin{equation}\label{Fcoll3}
{\bf q}_{coll}=-\nu_{eff}{\phi_s}^{k-1}\nabla\phi_s\,.
\end{equation}
From the constitutive viewpoint, the collision terms (\ref{Fcoll2}) or (\ref{Fcoll3})
can be considered as deriving from treating the ensemble of particles as a gas, 
so that the stress term for the solid constituent is isotropic through a coefficient that depends on the particle density.
Substituting Eqs. (\ref{Fcoll3}) and (\ref{closure}) into (\ref{mass})  gives the nonlinear degenerate advection-diffusion equation 
\begin{equation}\label{advdiff}
\frac{\partial \phi_s}{\partial t}+
\nabla\cdot(\phi_s\vv_f)-\dfrac{\partial}{\partial z}(\phi_s w_{sed})=
\nabla\cdot\left(\nu_{eff}\phi_s^{k-1}\nabla\phi_s \right)\,.
\end{equation}

Actually, if the limit value $k=1$ is also allowed, one has  the linear case sometimes used in the literature.

The coefficient $\nu_{eff}$ might be considered as composed of three contributions
$\nu_{eff}=\nu_s+\nu_{turb}+\nu_{mol}\,,$
that take into account of the possible dependence from 
\begin{itemize}
\item the shear rate, or better the velocity gradient $\nu_s=\nu_s(\nabla\phi_s)$;
\item the turbulence of the flow, so that this term results from the integration of the CFD simulation in a turbulent regime;
\item the molecular diffusion, but as reported in \cite{Kang12},  this term can be neglected with respect to other quantities. 
\end{itemize}

Actually, starting from the obvious observation that the behaviour of sand particles is isotropic. Objectivity, i.e., independence
of the constitutive dependence from the reference frame, implies that $\nu_s$ is a scalar isotropic function of a tensor.
By the representation theorem of isotropic function $\nu_s$ can then only depend on
the invariants of the rate of strain tensor $\D=\frac{1}{2}(\nabla\vv+\nabla\vv^T)$.
However, since the flow can be considered as a perturbation of a shear flow in the vertical plane, the leading contribution is the second invariant $I\!I_D=\frac{1}{2}\left[({\rm tr}\D)^2-{\rm tr}\D^2\right]$. For this reason we assume that 
$\nu_s=\nu_s(I\!I_D)$. This is an important generalization because all papers refers to a dependence on the wind velocity $u^*$, which is a well defined quantity only on horizontal surfaces, while more general surfaces like dunes require a dependence from an objective invariant of the rate of strain tensor.

In order to understand the meaning of the collision term we can look for stationary configurations for which all quantities depend only on the quote $z$ and velocities are directed along a flat plane (at $z=0$ in the direction $x$). In this case, the stationary profile in the saltation layer can be obtained integrating
$$\nu_{eff}\phi_s^{k-1}\frac{\partial \phi_s}{\partial z}+w_{sed}\phi_s =0\,,$$
jointly with the boundary condition at the sand bed.
In the simplest case in which $\nu_{eff}$ is constant and $k=1$, one immediately has an exponential profile with a characteristic length related to the thickness of the saltation layer given by $\delta_s=\nu_{eff}/w_{sed}$ and an integration constant related to the erosion boundary condition.

In general, referring to Fig. \ref{figure_2}c the dependence of the coefficient on the distance from the sand bed shows a strong diffusion closer to the surface and a low diffusion at a distance close to the height of the saltation layer, while particles that escape from the saltation layer are again strongly mixed due to the increasing diffusion due to turbulence.

The general features of the erosion boundary condition can be inferred from experiments known for a long time that show that, generally speaking, erosion only occurs if the shear stress at the surface $\tau$ exceeds a threshold level $\tau_t$, 
or equivalently that $u^*=\sqrt{\tau/\hat{\rho_f}}$ exceeds a threshold level $u_t^*=\sqrt{\tau_t/\hat{\rho_f}}$ (see \cite{Kokrev} and referencces therein).
%\cite{andreotti_parte1, Bagnold_1941, Duran_2010, Gillette_1974, Ji, Kawamura_1951, Kerr, Lettau,Nalpanis, Owen, Tsoar, Sauermann01, Shao_1999, Ungar}, 
%andreotti_parte1, bbagnold1941; Duran et al., 2010; Gillette et al., 1974; Ji et al. 2004; Kawamura_1951, Kerr and Nigra, 1952; Lettau and Lettau, 1977; Owen, 1964; Nalpanis et al., 1993; Pye and Tsoar, 1990; Sauermann et al., 2001; Shao and Li, 1999; Ungar and Haff, 1987), 

Referring to the last notation, because it
is the one classically used in the literature (though from the numerical point of view what is computed is $\tau$ which is then compared to $\tau_t$) one has the flux boundary condition
$$q_s=-\nu_{eff}\phi_s^{k-1}\nabla\phi_s\cdot {\bf n}=\beta\left(u^{*^2}-u^{*^2}_t\right)_+\,,$$
where $(f)_+=\frac{(f)+|f|}{2}$
stands for the positive part of $f$.
Bagnold's formula \cite{Bagnold_1941}
$u^*_t=\sqrt{\frac{\hat{\rho}_s-\hat{\rho}_f}{\hat{\rho}_f}}gd$
is quite successful in determining the dependence of $u^*_t$ from the grain diameter. On the other hand, the quantification of the
parameter $\beta$ is not straightforward because most of the experiments measures the horizontal sand
flux parallel to the surface while for the boundary condition one would need some knowledge of the vertical flux perpendicular to the surface which is the one related to the ejection of sand grains from the sand
bed.
Very recently, on the basis of their experiments Ho et al. \cite{Ho_2011, Ho_2012, Ho_2014} proposed 
$\beta=A_H\hat{\rho}_f\sqrt{{d}/{g}}$,
so that the sand flux due to erosion takes the form
\begin{equation}\label{BC}
q_s=\frac{w\alpha\hat{\rho}_f}{gd} \beta\left(u^{*^2}-u^{*^2}_t\right)_+\,,
\end{equation}
where $w = 0.5\,m/s$ is the sand grain ejection vertical velocity evaluated experimentally \cite{Creyssels_2009, Ho_2011} and $\alpha$ is a dimensionless free parameter to be fitted to experimental sand flux profiles.
%
%%%%%%%%%%%%%%%%%%%%%%%%%%%%%%%%
\section{Simulations}
As domain of integration we focus on the flow over a horizontal heterogeneous lane. This is neither an artificial situation, nor a case of limited importance. In fact, most of the landforms in arid regions and roads are well approximated by a horizontal flat plane.
This geometrical setup is also retained in most of the wind tunnel experimental studies present in the literature that are however mainly addressed to the characterisation of the windblown sand concentration and flux in uniform, in-equilibrium steady state conditions. 
Nevertheless, uniform and steady state conditions are excessively ideal ones. In these situations
the incoming wind already transports the maximum allowable sand density, so that erosion and deposition are balanced, and hence the sand bed surface is neither scoured, nor accumulated. 
\begin{figure}[h!]
\begin{center}
\includegraphics[width=.7\textwidth]{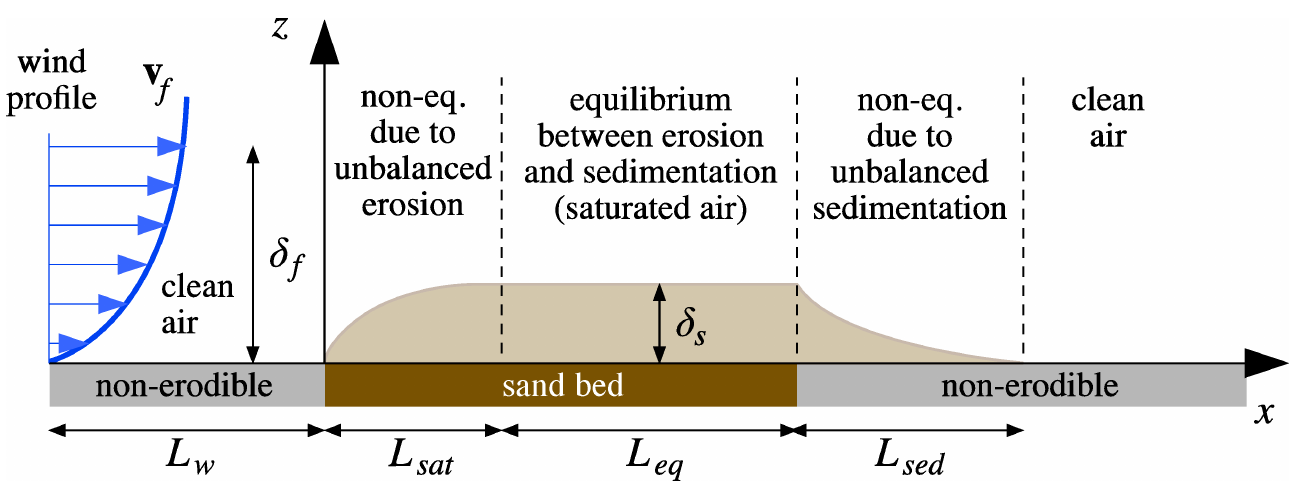}
\end{center}
\vspace{-.8cm}
\caption{
Considered setup with alternation of erodible and non-erodible surfaces, of equilibrium and non-equilibrium regions, and a sketch of the height of the saltation layer $\delta_s$.}
\label{figure_3}
\end{figure}
Conversely, in many engineering applications, attention must be paid to non-equilibrium conditions that will cause erosion or settlement of the sand bed. So, as sketched in Fig. \ref{figure_3}, our horizontal plane is
characterized by the alternation of erodible sandy regions and non-erodible regions, corresponding for instance to a street.
In the simulation the inflow wind is clean and with a fully developed logarithmically shaped velocity profile, with $z_0= 10^{-5} \,m$  and $u^*$ ranging between 0.3 and 1 $m/s$. The threshold value for erosion is $u^*_t=0.25	\,m/s$ and the grain size is $d=0.25 \,mm$.

\begin{figure}[h!]
\begin{center}
\includegraphics[width=.8\textwidth]{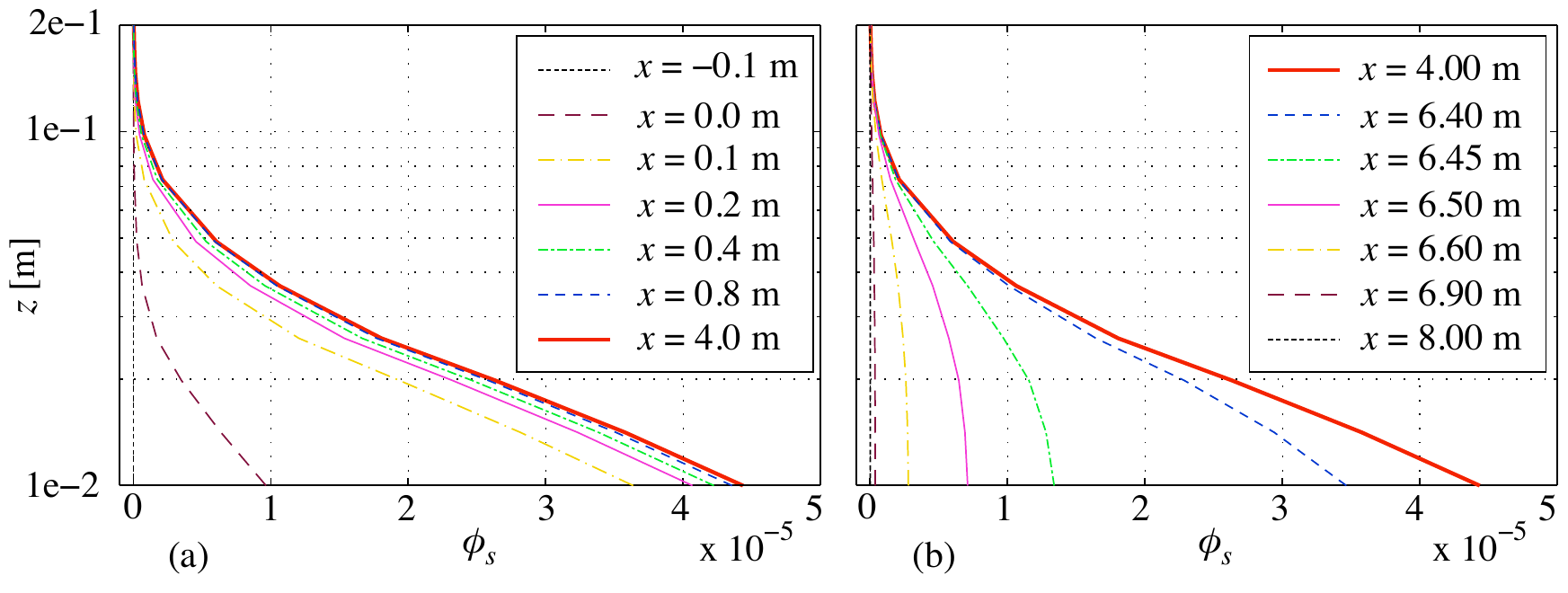}
\end{center}
\vspace{-.8cm}
\caption{
Sand density profiles in the saturation (a) and in the sedimentation (b) regions}
\label{figure_4}
\end{figure}

\begin{figure}[h!]
\begin{center}
\includegraphics[width=.8\columnwidth]{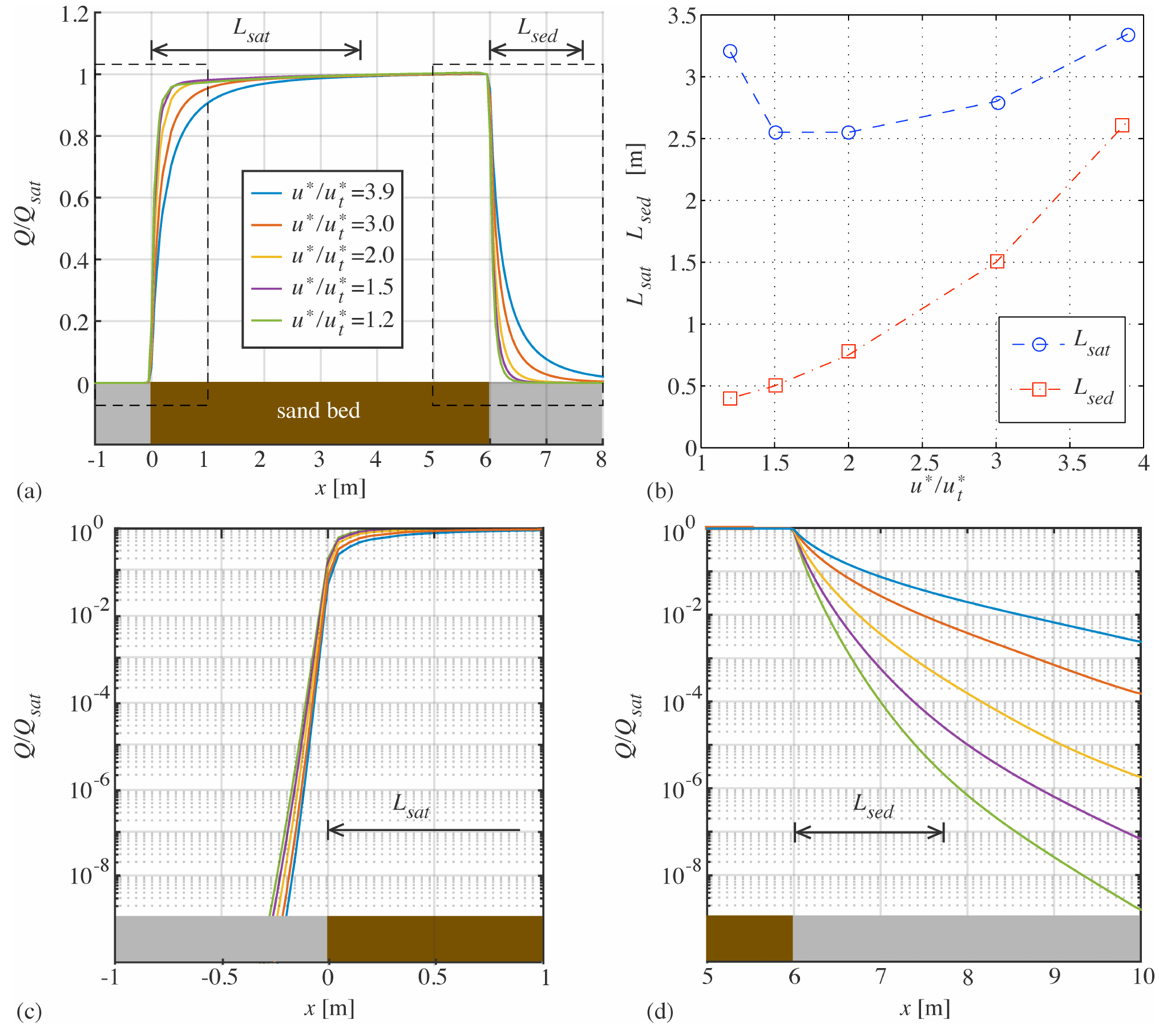}
\end{center}
\vspace{-.8cm}
\caption{
Scaled total sand flux over the sand bed at different wind velocities}
\label{figure_5}
\end{figure}

From the simulation shown in Figures \ref{figure_4} and \ref{figure_5}, as soon as wind overcomes the boundary between the non-erodible and erodible surface (that is put at a distance   $L_w=1\,m$ from the inflow boundary) the saltation layer starts to develop. Figure \ref{figure_4}a plots the profiles of the sand volume ratio $\phi_s$ at several points at the beginning of the erodible zone. The model correctly predicts the progressive uptake of sand and increase in the depth of the saltation layer, till saturation is reached because of  the  equilibrium between erosion and sedimentation. The thickness of the saltation layer at equilibrium is about 10 cm in qualitative agreement with experiments. Viceversa, as shown in Fig. \ref{figure_4}b, at the beginning of the second non-erodible zone, there
is a reduction of the windblown sand density in the upper part of the stream because of the sedimentation process, not balanced by saltation.

Actually, due to diffusion, some sand also diffuses upstream, mainly very close to the surface where diffusion is dominant. Referring to Fig. \ref{figure_5}c, in this region one can notice an exponential growth of the scaled total sand flux $Q/Q_{sat}$ where $Q_{sat}$ is the sand flux at equilibrium. After the first soil discontinuity at $x=0$ convection dominates and $Q$ saturates in a length close to $L_{sat}=3\,m$ (see Fig. \ref{figure_5}b), that is nearly independent from the scaled wind velocity $u^*/u^*_t$, in qualitative agreement with \cite{Andreotti_2010}. 

When the erodible surface ends the total sand flux $Q $ readily decreases in a characteristic distance that increases with the wind velocity as shown in Fig. \ref{figure_5}b. It can be noticed from Fig. \ref{figure_5}d that the decrease is less than exponential in qualitative agreement with \cite{Ho_2014}.

In conclusion, the model (\ref{NS},\ref{advdiff}) jointly with the fundamental erosion boundary condition (\ref{BC}) and other standard initial and boundary conditions is able not only to describe the erosion/transport/sedimentation process in stationary situation, but also to capture all features characterizing the development of the saltation layer up to equilibrium and of its reduction, that occur in nature when the sandy surfaces is heterogeneous. 

\end{document}